\documentclass[twocolumn, astrosym, appendixfloats]{aastex6}
\usepackage{natbib}
\usepackage{amsmath}
\usepackage{amssymb}
\usepackage{xspace}
\usepackage{graphicx,etoolbox}

\usepackage{hyperref}
\hypersetup{
    colorlinks = true,
    linkcolor = blue,
    citecolor = blue,
    urlcolor = magenta
}

\graphicspath{{./}}

\newcommand{\figsize}{0.48\textwidth}
\defcitealias{Hasani2016}{Paper I}
\defcitealias{Marks2012}{Marks et al}

\newcommand{\Msun}{\,\mathrm{M}_{\sun}}
\newcommand{\pc}{\,\mathrm{pc}}
\newcommand{\FeH}{\mathrm{[Fe/H]}}
\newcommand{\Mc}{M_\mathrm{c}}
\newcommand{\Gyr}{\,\mathrm{Gyr}}
\newcommand{\Myr}{\,\mathrm{Myr}}
\newcommand{\Nbody}{$N$-body\xspace}
\newcommand{\secref}[1]{Section~\ref{#1}}
\newcommand{\figref}[1]{Figure~\ref{#1}}
\newcommand{\tabref}[1]{Table~\ref{#1}}
\newcommand{\equref}[1]{equation~\eqref{#1}}

\newcommand{\twopartdef}[4]
{
    \left\{
        \begin{array}{ll}
            #1 & \mbox{ \ } #2 \\
            #3 & \mbox{ \ } #4
        \end{array}
    \right.
}

\begin{document}

\title{A Possible Solution for the $M/L-\FeH$ Relation of Globular Clusters in M31. \\ II. the age-metallicity relation}
\shorttitle{Solution for $M/L$ ratios of GCs in M31}
\shortauthors{Haghi et al.}

\author{Hosein Haghi\altaffilmark{1},
	Pouria Khalaj\altaffilmark{2},
	Akram Hasani Zonoozi\altaffilmark{3}}
\affil{Institute for Advanced Studies in Basic Sciences (IASBS), P.O. Box 45195-1159, Zanjan, Iran}

\and
\author{Pavel Kroupa\altaffilmark{4}}
\affil{Helmholtz-Institut f$\rm\ddot{u}$r Strahlen- und Kernphysik (HISKP), University of Bonn, Nussallee 14-16, 53115 Bonn, Germany}

\altaffiltext{1}{haghi@iasbs.ac.ir}
\altaffiltext{2}{pouria.khalaj@uqconnect.edu.au; School of Mathematics and Physics, University of Queensland, St. Lucia, QLD 4072, Australia}
\altaffiltext{3}{a.hasani@iasbs.ac.ir}
\altaffiltext{4}{pavel@astro.uni-bonn.de; Charles University in Prague, Faculty of Mathematics and Physics, Astronomical Institute, V Hole\v{s}ovi\v{c}k\'ach 2, CZ-180 00 Praha 8, Czech Republic}

\begin{abstract}
This is the second of a series of papers in which we present a new solution to reconcile the prediction of single stellar population (SSP) models with the observed stellar mass-to-light ($M/L$) ratios of globular clusters (GCs) in M31 and its trend with respect to $\FeH$. In the present work our focus is on the empirical relation between age and metallicity for GCs and its effect on the $M/L$ ratio. Assuming that there is an anti-correlation between the age of M31 GCs and their metallicity, we evolve dynamical SSP models of GCs to establish a relation between the $M/L$ ratio (in the $V$ and $K$ band) and metallicity. We then demonstrate that the established $M/L-\FeH$ relation is in perfect agreement with that of M31 GCs. In our models we consider both the canonical initial mass function (IMF) and the top-heavy IMF depending on cluster birth density and metallicity as derived independently from Galactic GCs and ultra-compact dwarf galaxies by \citetalias{Marks2012}. Our results signify that the combination of the density- and metallicity-dependent top-heavy IMF, the anti-correlation between age and metallicity, stellar evolution and standard dynamical evolution yields the best possible agreement with the observed trend of $M/L-\FeH$ for M31 GCs.
\end{abstract}



\section{Introduction}\label{sec:introduction}

The $M/L$ ratio is a vital tool for studying star clusters and can be used as a diagnostic to constrain e.g. the IMF and the age of the cluster. In a GC there are two opposing mechanisms which affect the $M/L$ ratio, namely stellar evolution and dynamical evolution. As the cluster ages, high-mass stars ($m>1\Msun$) which have a faster evolution rate and account for the bulk of the luminosity of the cluster turn into compact remnants with high $M/L$ ratios. This increases the $M/L$ ratio with time. On the other hand, as the cluster tends towards energy equipartition the low-mass stars which constitute a large fraction of the cluster mass are preferentially lost as a direct consequence of two-body relaxation. This reduces the $M/L$ ratio with time. \citet{Baumgardt2003} performed a series of \Nbody simulations of GCs in external tidal fields. They adopted a canonical IMF \citep{Kroupa2001} for their model clusters assuming that GCs are comprised of a single stellar population (SSP). They showed that the effect of dynamical evolution on reducing $M/L$ ratios becomes especially important for clusters in strong tidal fields and in advanced evolutionary phases, i.e clusters which have lost $60\%$ or more of their mass.

\par \citet{Strader2009, Strader2011} derived the structural properties, kinematical properties and the $M/L$ ratios of 163 GCs in the M31 galaxy in the near infrared ($K$-band) and optical ($V$-band). Their sample of GCs exhibit $M/L$ ratios which are considerably lower than those predicted from SSP models of GCs with a canonical IMF. In addition, one expects that the $M/L$ ratios derived from SSP models to show a positive correlation with metallicity. This is, however, at odds with the observations of M31 GCs as their $M/L$ ratios show an inverse trend. For example, at solar-metallicity ($\FeH=0$) the $M/L$ ratios of M31 GCs in the $V$ band are lower than those of SSP models (with $T=12.5\Gyr$) by a factor of 3. Similar results have been found for Galactic GCs (GGCs) as well \citep{Kimming2015}.

\par A number of studies have addressed this discrepancy and proposed solutions which are mainly based on the depletion of low-mass stars either due to dynamical evolution \citep{Kruijssen2009} or a bottom-light IMF \citep{Strader2011}. However, \citet{Shanahan2015} examined the bias that the assumption of \textit{light follows mass} would introduce in the determination of the $M/L$ ratios of mass-segregated GCs. In particular, they quantified the effect of mass segregation on the perceived $M/L$ ratio of GCs as a function of $\FeH$. GCs with higher metallicities are observed to have smaller projected half-light radii compared to GCs with the same age but lower metallicities. This is due to the fact that the turn-off mass of a cluster tends towards a larger value with increasing metallicity, implying that the cluster has more bright (massive) stars and they are more centrally concentrated (due to mass-segregation) in GCs with higher metallicities. As a result, if a GC is unresolved and multi-mass models cannot be fit to determine its properties, then the combination of the metallicity-dependent mass-segregation and the assumption that light follows mass will lead to a bias in the determination of $M/L$ ratio. \citet{Shanahan2015} showed that such a bias will lead to an underestimation of the cluster mass. As a result, if one accounts for the bias that exists in the inferred $M/L$ ratios of GCs from their integrated light properties, the predictions of SSP models are compatible with observations, hence assuming anomalous IMFs for clusters is not required according to their findings. \citet{Baumgardt2016} determined the $M/L$ ratios of 50 well observed GCs using \Nbody simulations. They showed that the $M/L$ ratios of their studied GCs are compatible with a standard \citet{Kroupa2001} or \citet{Chabrier2003} IMF, and except for $\FeH\gtrsim-1$ where observed $M/L$ ratios are $\approx20\%$ lower than what is predicted by simulations, they did not find any evidence for a decrease of the $M/L$ ratios with metallicity. However, they stated that their findings are not conclusive as more accurate $M/L$ ratios or a wider range of cluster parameters in their simulations are needed.

\par More recently, \citet{Hasani2016}, hereafter \citetalias{Hasani2016}, showed that a metallicity- and density-dependent top-heavy IMF \citep{Marks2012} in conjunction with dynamical evolution is able to successfully explain both the observed low $M/L$ ratios of M31 GCs in the $K$ band and their trend with respect to metallicity. In the $V$ band, however, there was a minor discrepancy between our results and observations.
More precisely, despite the fact that we were able to decrease the $M/L_{V}$ ratios of SSP models significantly and make them closer to the observed values of M31 GCs, the observed $M/L_V$ ratios were still lower than those of our models. In the present paper which is a follow-up to \citetalias{Hasani2016}, we propose a solution to resolve this discrepancy and reach an even better agreement with observations both in the $V$ and $K$ band. Our solution is based on the relation between the age and the metallicity of the GCs, hereafter the age-metallicity relation (AMR).

\par The paper is structured as follows. In the next section we discuss the AMR and its effect on the $M/L$ ratio of a GC. We summarize and conclude our work in \secref{sec:conclusion}.


\begin{figure*}[t!]
	\centering
	\includegraphics[width=\figsize]{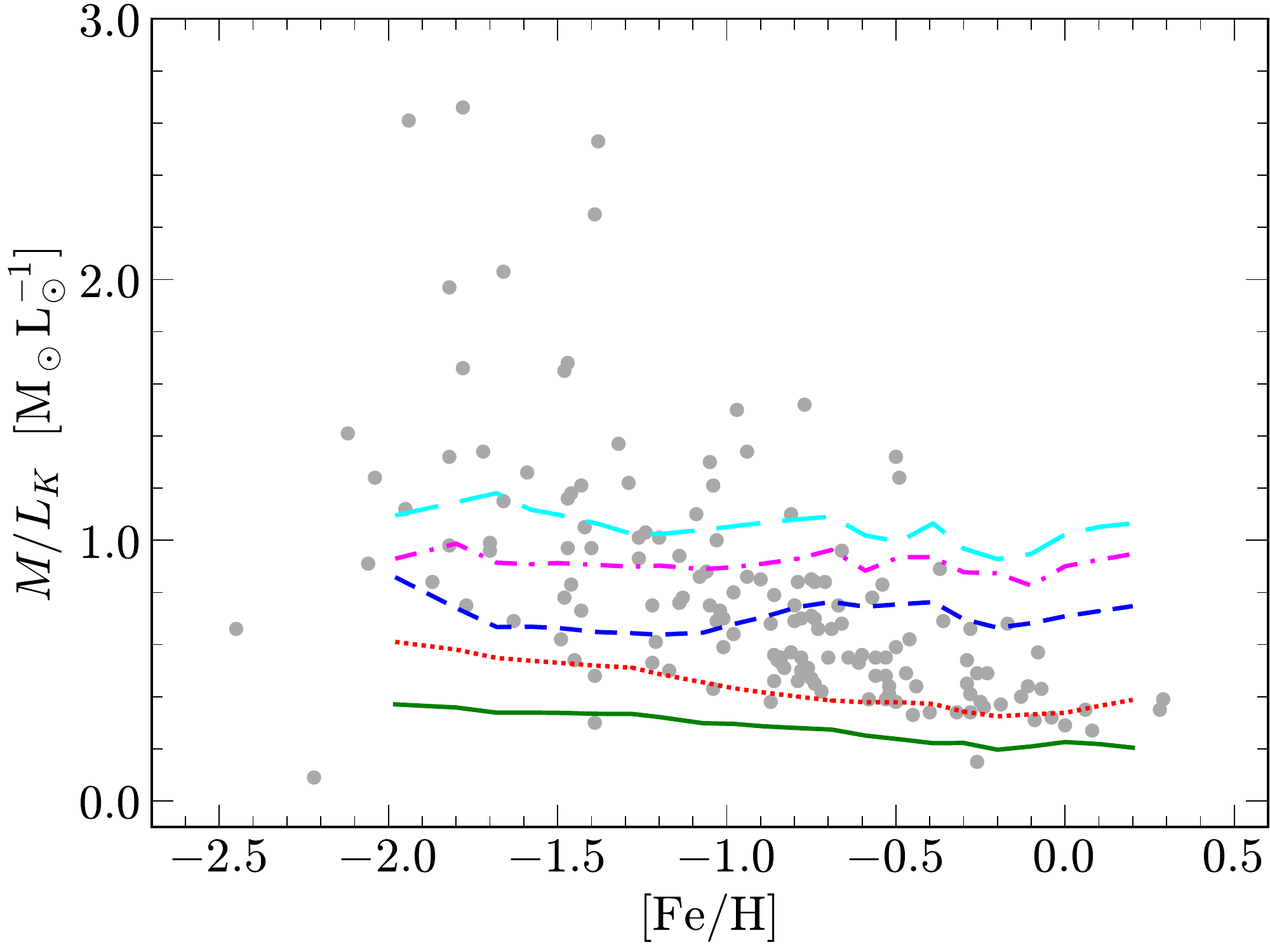}
	\includegraphics[width=\figsize]{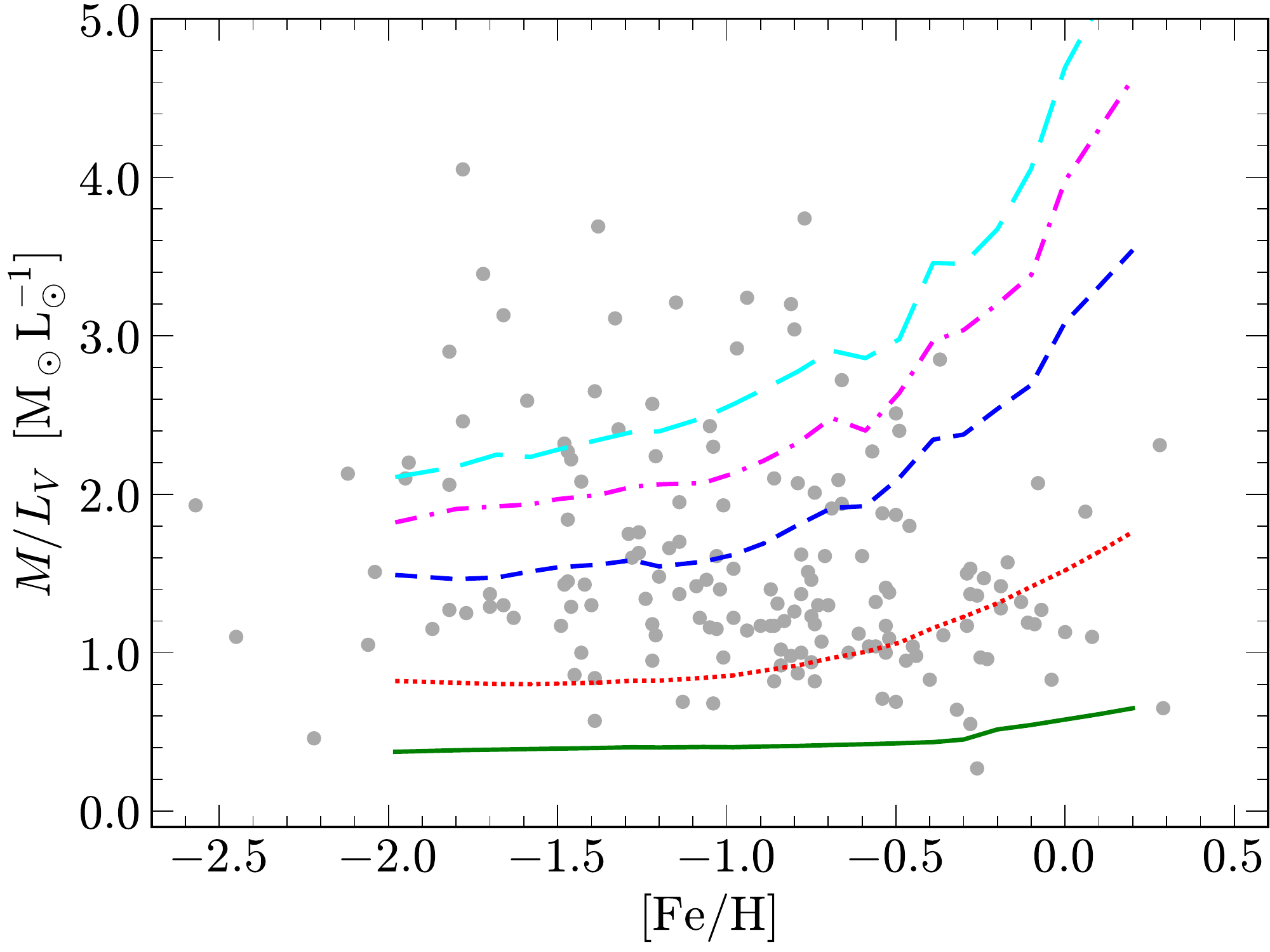}	
	\caption{The $M/L$ ratios vs. [Fe/H] for M31 GCs (denoted by grey dots). Data is taken from \citet{Strader2011}. The left and the right panel show the $M/L$ ratios in the $K$ and $V$ band respectively. The coloured lines show the prediction of SSP models with a canonical IMF \citep{Kroupa2001} but different ages and correspond to $T=1\Gyr$ (solid green line), $T=3\Gyr$ (dotted red line), $T=7\Gyr$ (dashed blue line), $T=10\Gyr$ (dash-dotted magenta line) and $T=12.5\Gyr$ (long-dashed cyan line). All clusters have the same initial mass of $\Mc=10^6\Msun$. The SSP curves in this plot do not include the effect of dynamical evolution.}
	\label{fig:SSP}
\end{figure*}

\begin{figure*}[t!]
  \centering
    \includegraphics[width=\figsize]{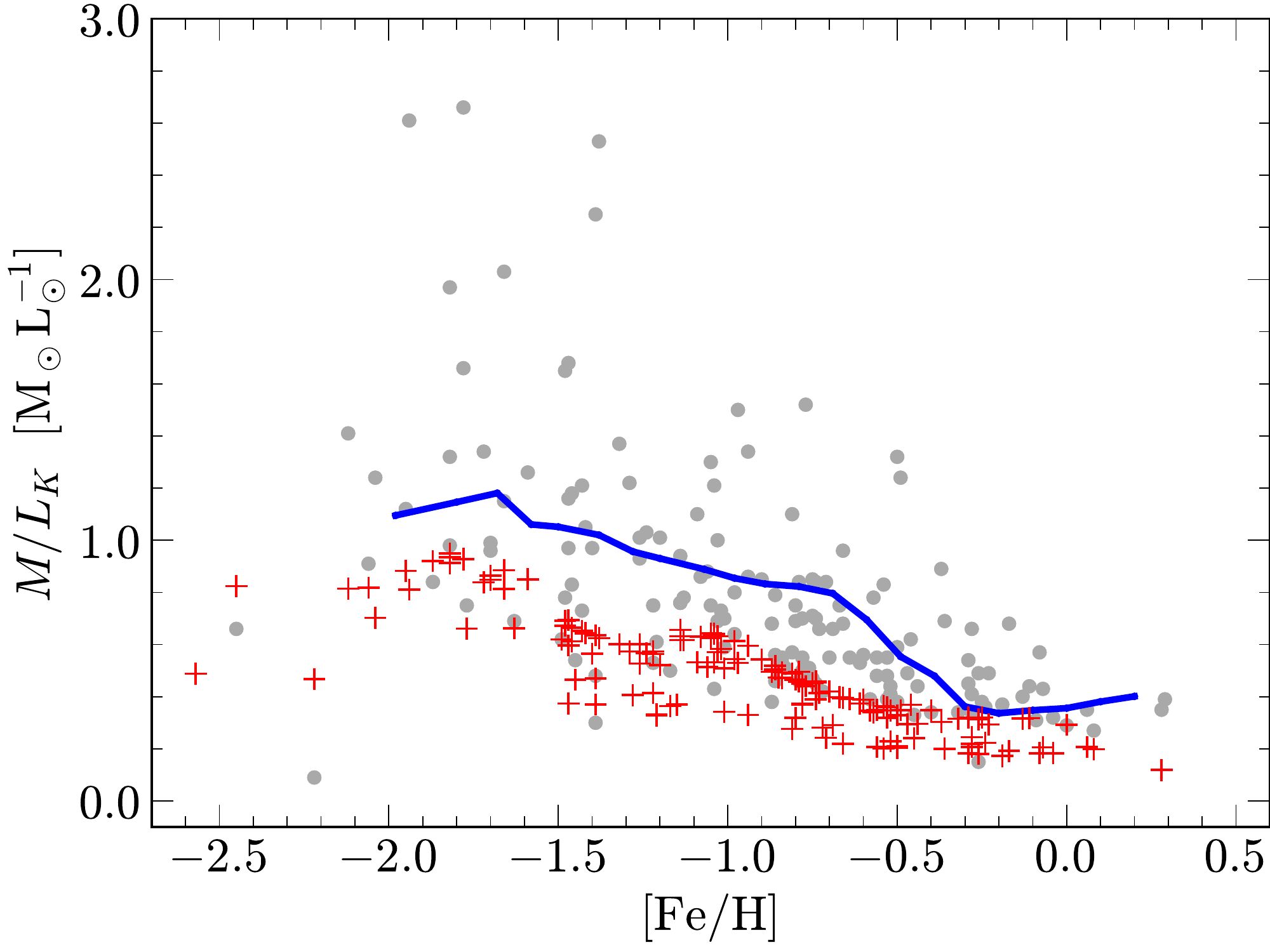}
    \includegraphics[width=\figsize]{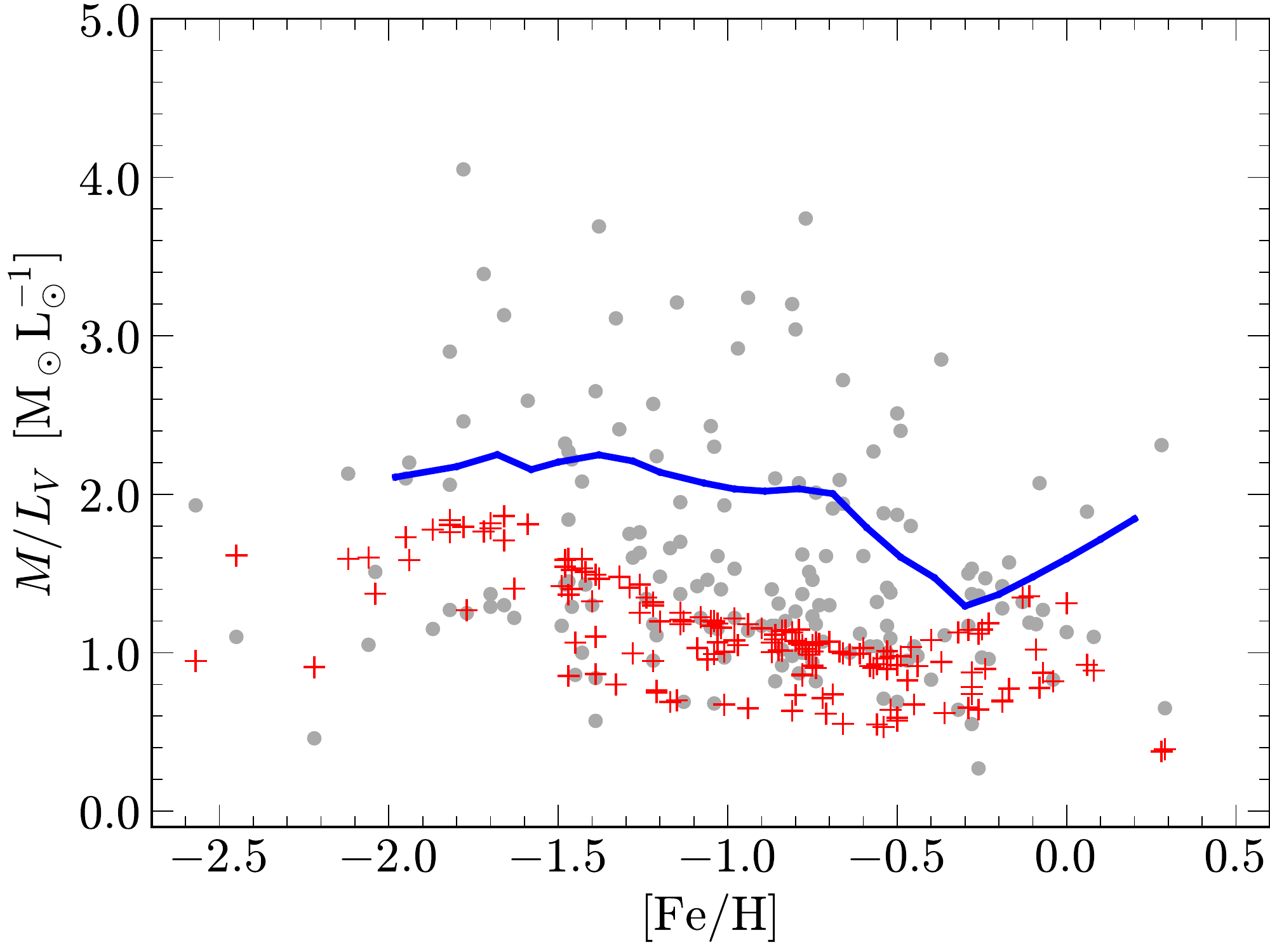}
    \caption{The $M/L$ ratios of SSP models with a canonical IMF (denoted by red crosses) calculated at the $\FeH$ values of M31 GCs (grey dots) considering the AMR (solid blue line) from \citet{Cezario2013} valid for M31 GCs and dynamical evolution. See the text for more details. The WD retention fraction is $100\%$.}
    \label{fig:scatter-canonical}
\end{figure*}

\begin{figure*}[t!]
  \centering
  \includegraphics[width=\figsize]{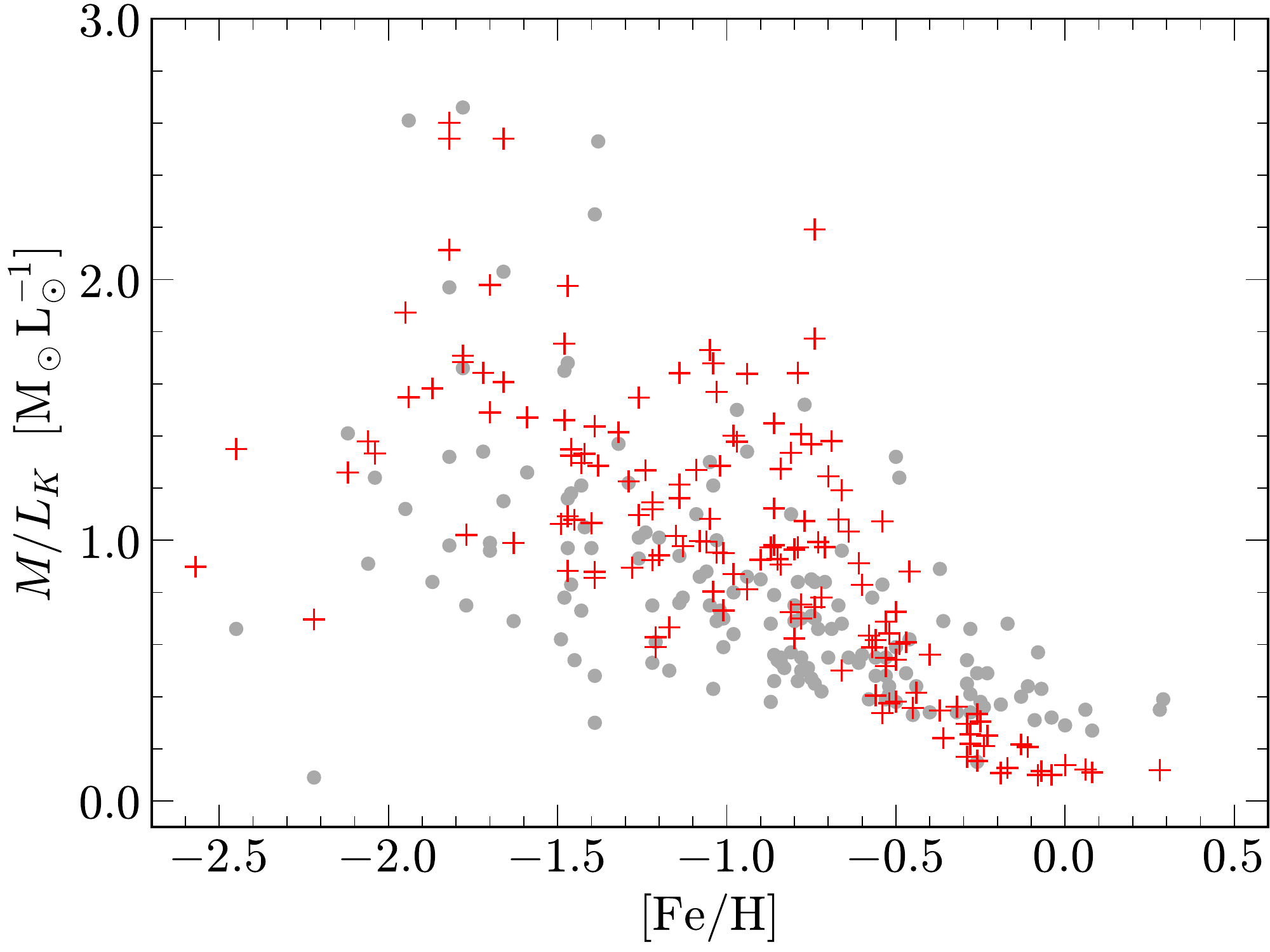}
  \includegraphics[width=\figsize]{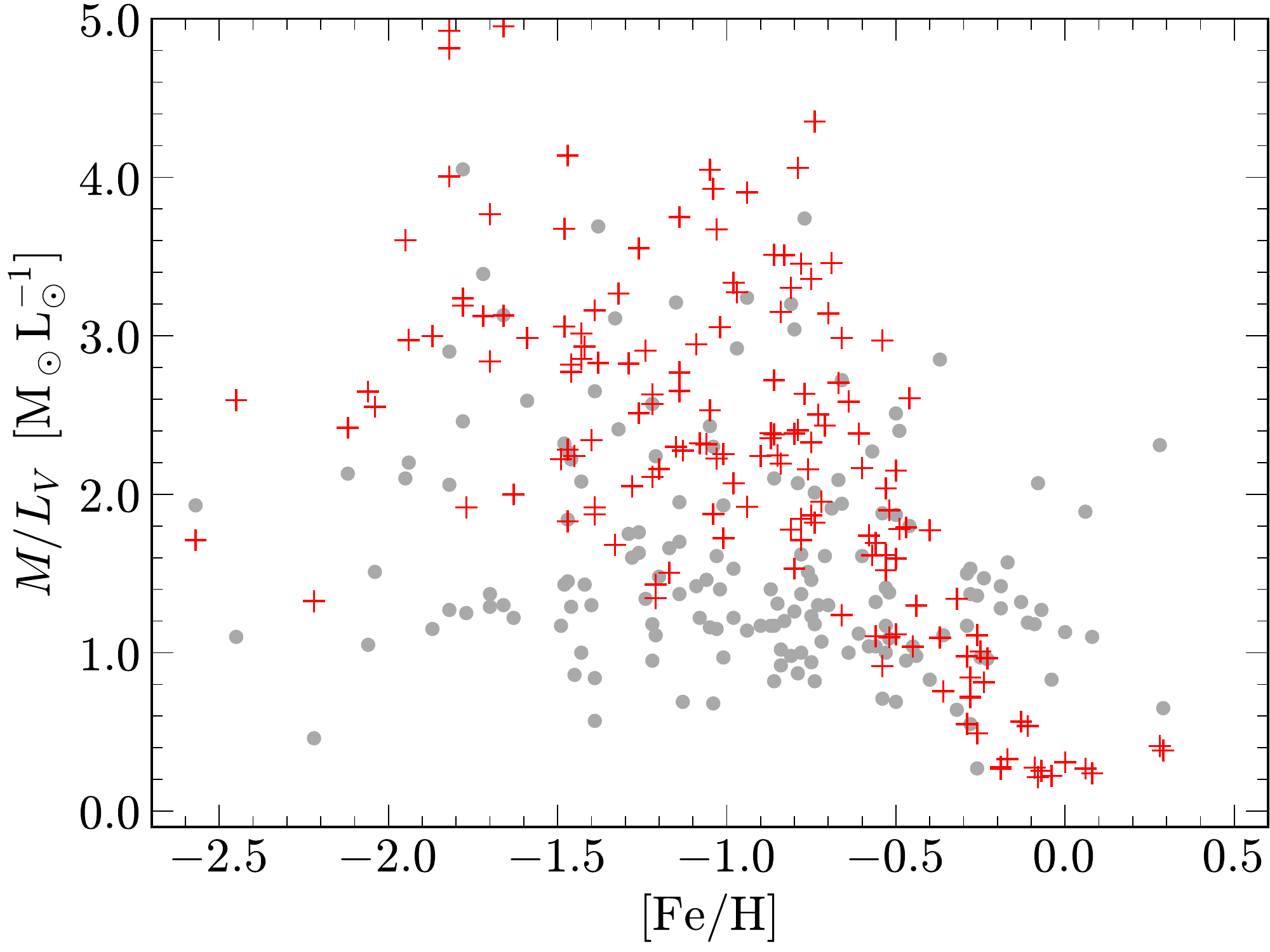}
  \includegraphics[width=\figsize]{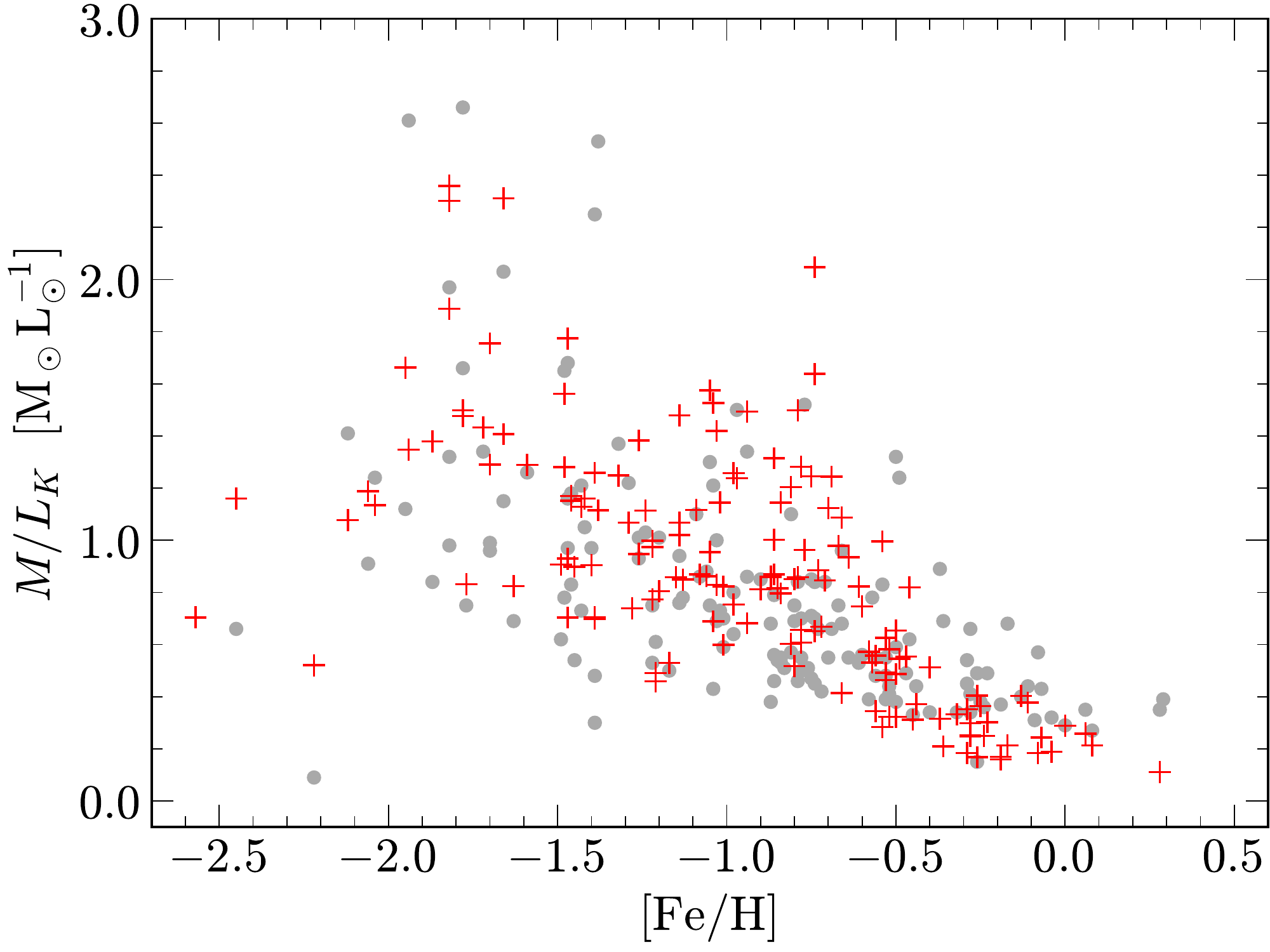}
  \includegraphics[width=\figsize]{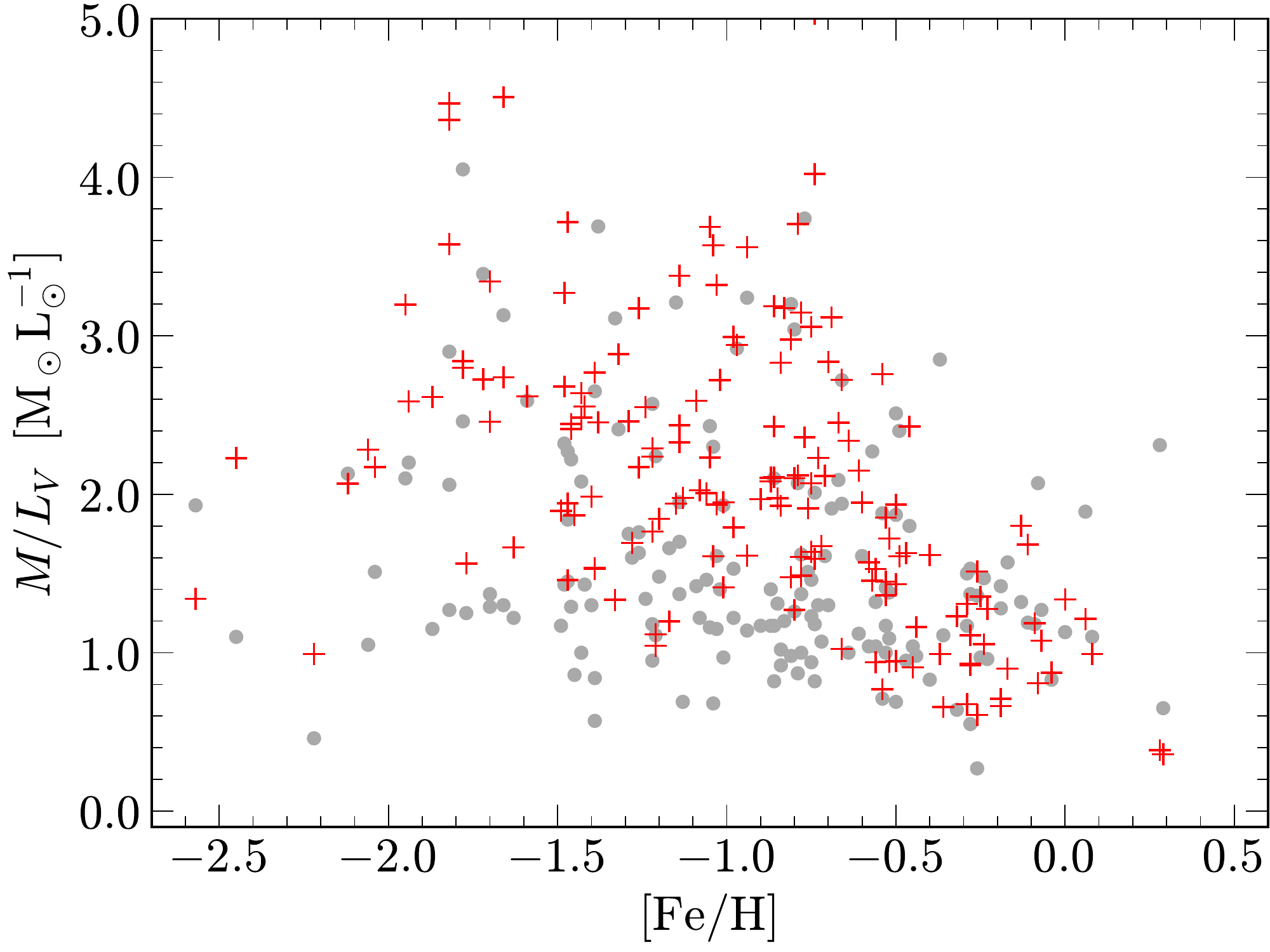}
  \caption{Same as \figref{fig:scatter-canonical} but with the addition of a metallicity- and density-dependent IMF (Eq.~\ref{eq:alpha3}). The WD retention fraction is 100\% and 50\%, respectively, for the top and bottom row. The NS and BH retention fractions depend on the initial mass and the half-mass radius of the clusters. The AMR used in this figure is from \citet{Salaris2002} being valid for the Milky Way. \newline\newline}\label{fig:scatter-topheavy-MW}
\end{figure*}

\begin{figure*}[t!]
  \centering
  \includegraphics[width=\figsize]{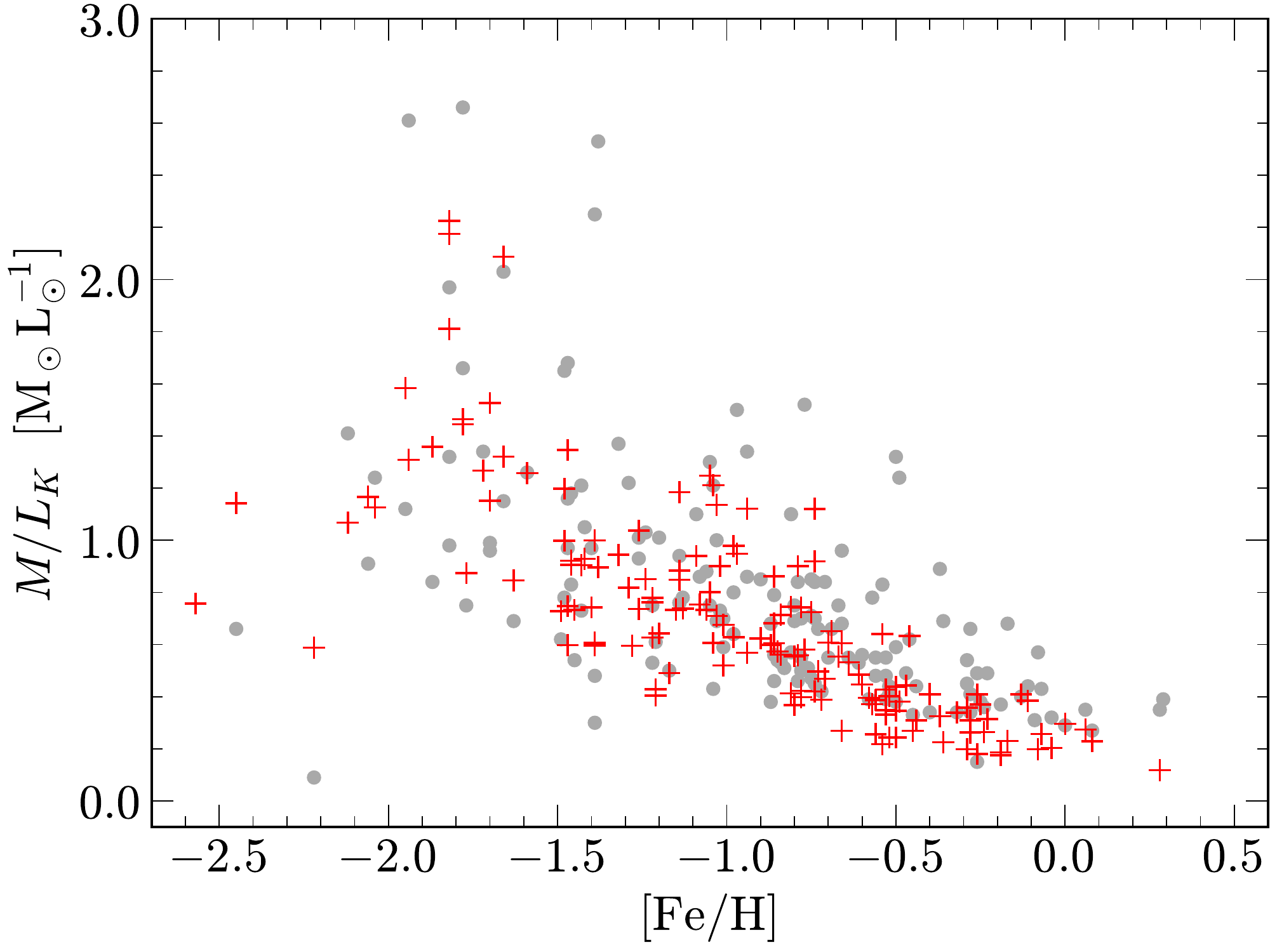}
  \includegraphics[width=\figsize]{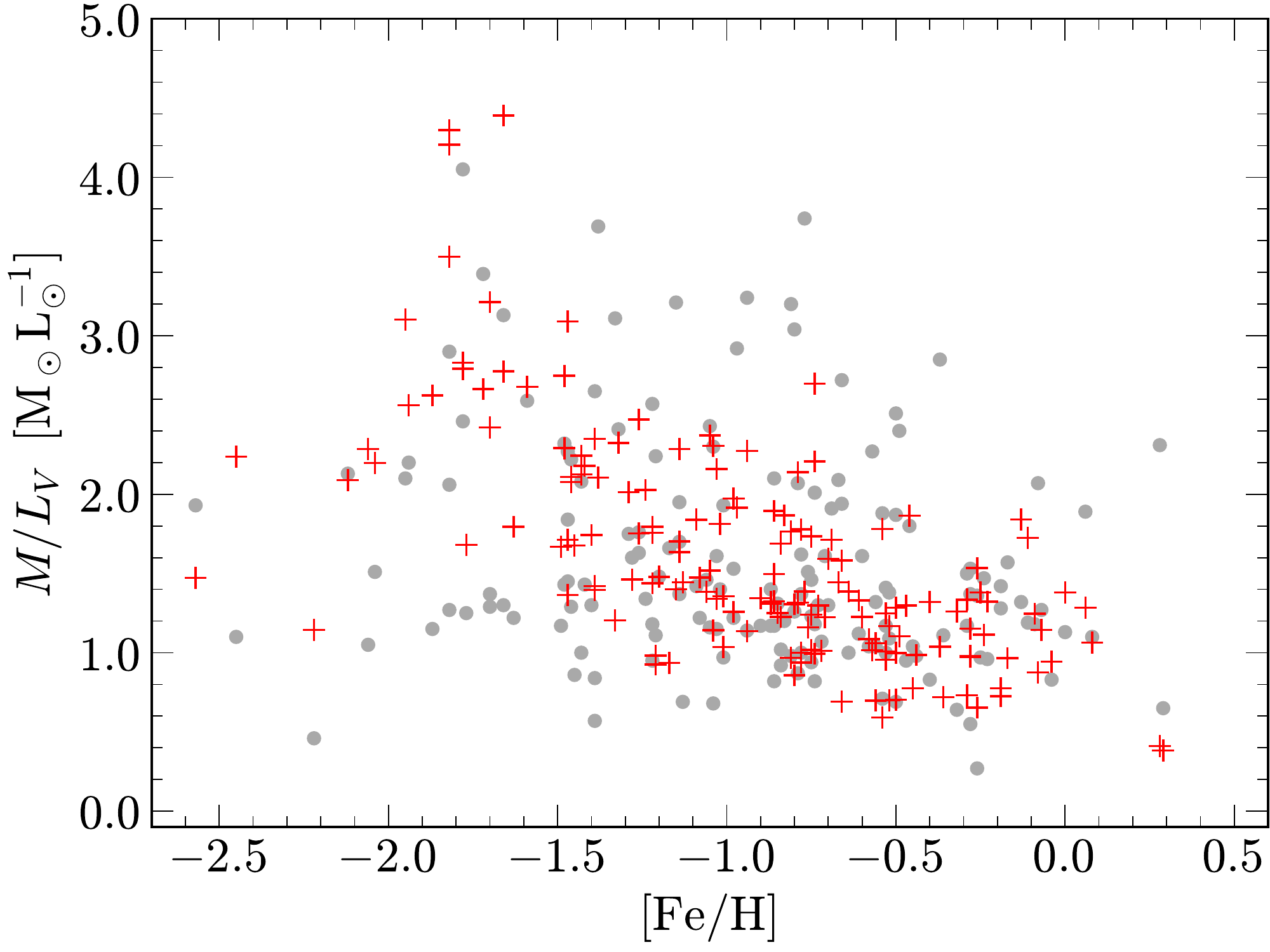}
  \includegraphics[width=\figsize]{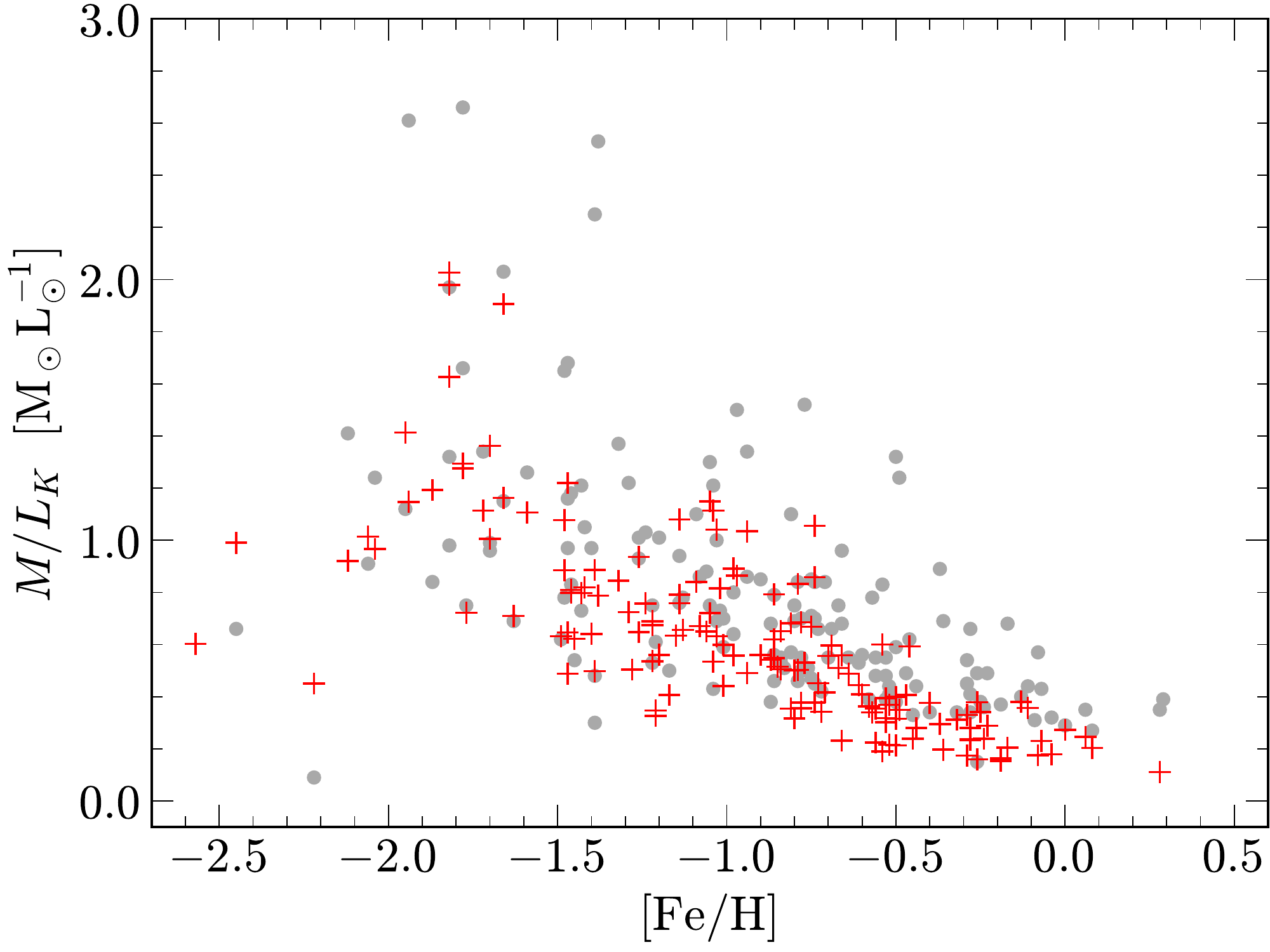}
  \includegraphics[width=\figsize]{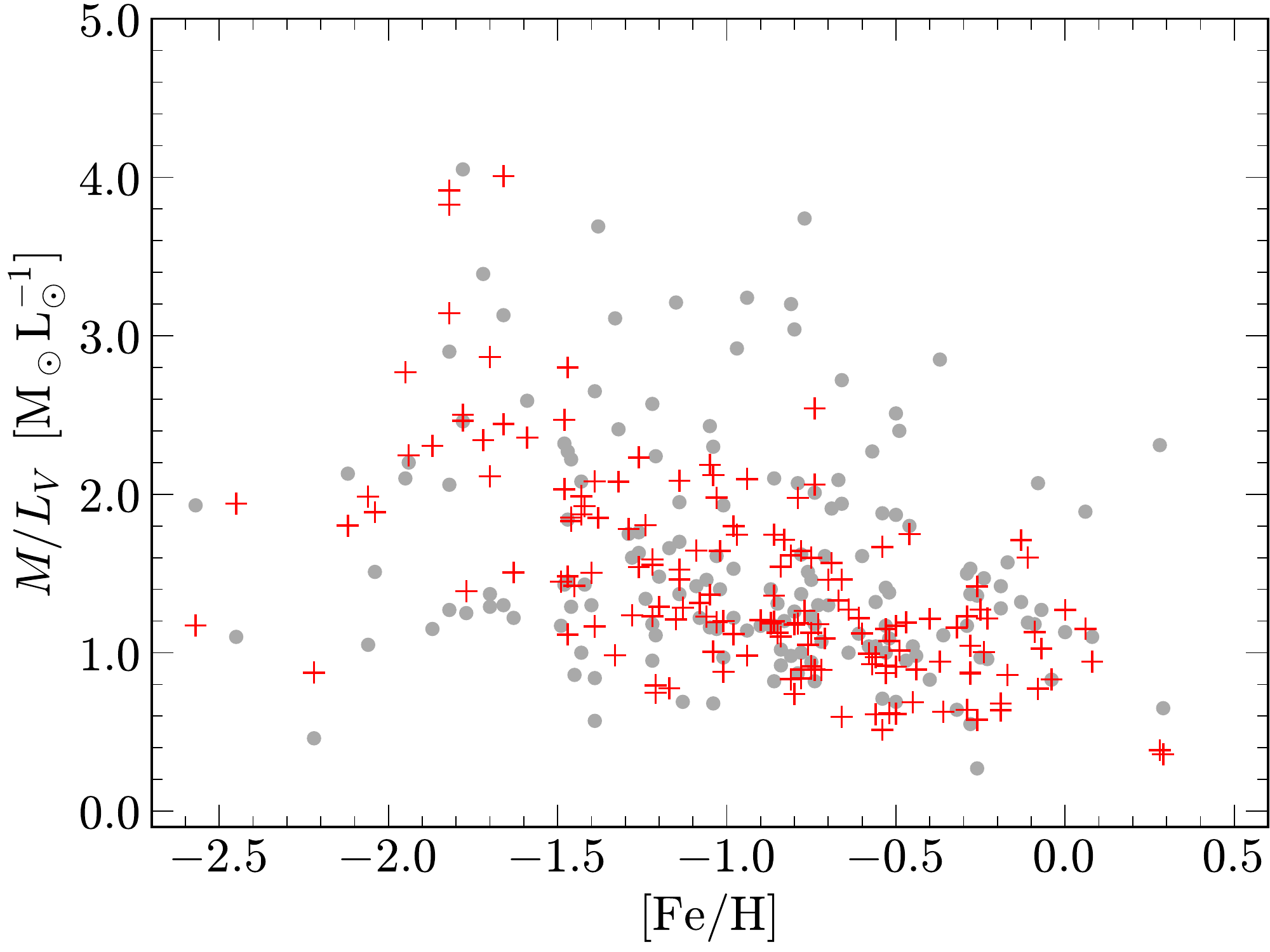}
  \caption{Same as \figref{fig:scatter-topheavy-MW} but for the AMR for M31 from \citet{Cezario2013}. \newline\newline}\label{fig:scatter-topheavy-And}
\end{figure*}

\section{AMR and the $M/L$ ratios of GCs}\label{sec:IMFAMR}

\par Studies of GCs have shown a variety of AMRs in the past, e.g. the Milky Way \citep{Salaris2002, Mendel2007, VandenBerg2013, Leaman2013, Roediger2014}, M31 \citep{Jiang2003, Fan2006, Cezario2013}, the Large Magellanic Cloud \citep{Carrera2011}, and NGC~147, NGC~185 and NGC~205 \citep{Sharina2006}. What is common to all of the AMRs found by these studies is the anti-correlation between age and metallicity, which is expected as the metallicity of GCs reflects the metallicity of the environment in which they have formed. Older GCs have formed early on when galaxies were still forming and when the interstellar medium was not strongly polluted by metal-rich material (e.g. winds of massive and rapidly evolving stars). In contrast, younger GCs have formed when the galaxies were already in place. This is supported by different spatial distributions of metal-poor and metal-rich GCs in elliptical (e.g. \citealt{Forbes2011}) and spiral galaxies (e.g. \citealt{Griffen2010}).

\par What differs in the studies of the AMR is the degree to which the inferred age depends on the measured metallicity (and vice versa), and also a possible dichotomy or bifurcation in the observed AMR (e.g. \citealt{Leaman2013}), which signifies the existence of two components of GCs, i.e. metal-poor and metal-rich, formed in different environments.

\par The existence of the AMR dictates that in a sample of randomly selected GCs, metal-poor GCs have a larger age on average. The dependence of $M/L_V$ and $M/L_K$ ratios on age, metallicity and initial cluster mass is nontrivial because SSP $M/L$ ratios increase with age, but decrease due to dynamical evolution of the clusters and are larger for metal-rich SSPs. In addition, the IMF plays a role by providing stellar remnants and low mass stars with large $M/L$ ratios.
To quantify this effect, we calculate SSP models, including GC dynamical evolution and proceed as follows.

\par To be consistent with \citetalias{Hasani2016}, the setup of our models and the recipe for stellar and dynamical evolution are the same as the ones that we used for \citetalias{Hasani2016}. First we consider the case of a canonical IMF. Using the flexible stellar population synthesis code (FPS, \citealt{Marigo2007, Marigo2008, Conroy2009, Conroy2010a, Conroy2010b}), we evolve GCs with an initial mass of $\Mc=10^6\Msun$ and a canonical IMF which extends from $0.08\Msun$ to $100\Msun$. \figref{fig:SSP} compares the distribution of M31 GCs in the $M/L-\FeH$ plane and the yields of our SSP models for ages ranging from $T=1\Gyr$ to $T=12.5\Gyr$. These models do not include the effect of dynamical evolution yet. One can see that the $M/L$ ratio of SSP models increases with age. In addition, in the $K$ band the $M/L$ curves are almost flat whereas in the optical band they strongly rise at the metal-rich end of the sample.

\par We take the age and metallicity data of a sample of 38 GCs in M31 from \citet{Cezario2013} which has been obtained by using a spectral fitting technique to the observed integrated spectra of GCs. This sample has the benefit of covering a wide range of age and metallicities, e.g. GCs as young as $150\Myr$. We fit a two-component and continuous mathematical function of the following form to the data presented in Table~3 of \citet{Cezario2013} to establish the AMR in M31,
\begin{equation}\label{eq:AMRM31}
\log_{10}(T/\Gyr) = \twopartdef {0.985}{\FeH \leq -1.6,}{a\FeH+b}{\rm{otherwise,}}
\end{equation}
where $T$ is the age of GCs in Gyr and $a$ and $b$ are two constants. The value of $\log_{10}(T/\Gyr)=0.985$ corresponds to the average age of old GCs in M31. The range of metallicity and age for the \citet{Cezario2013} sample that we have used to obtain the AMR for M31 GCs is $\FeH\in[-2.3, +0.1]$ and $T\in[0.2\Gyr, 13.0\Gyr]$ respectively. For comparison and in order to see how sensitive our results are to the adopted AMR, we fit a linear function ($\log_{10}(T/\Gyr)=a\FeH+b$) to the age and metallicity data of GGCs taken from \citet{Salaris2002}. For M31 GCs and GGCs, we find $(a, b)$ to be $(-0.46, 0.25)$ and $(-0.37, 0.51)$ respectively.\par Next, we incorporate the derived AMR with our SSP models. To do so, we take the metallicity of each GC in M31 and then use \equref{eq:AMRM31} to find its age using (a,b) for M31. Having found the age we then proceed to evolve the SSP model of the GC up to the specified age and find the $M/L$ ratio at that age.This enables us to establish a relation between the metallicity and the $M/L$ ratio of a cluster. The result is depicted by the solid blue line in \figref{fig:scatter-canonical}. As was expected, the introduction of the AMR into our models leads to an anti-correlation between $M/L$ and $\FeH$ which resembles the observed trend of M31 GCs. This is intriguing as even without considering the effect of dynamical evolution or a top-heavy IMF, the use of the AMR leads to a remarkable agreement with observations. This result is independent of the assumed mathematical form for the AMR. One can adopt any other mathematical form, e.g. a polynomial, as long as it captures the general trend of age and metallicity.

\par To account for the effect of dynamical evolution in our study, we take our calculated SSP models with either the canonical or the metallicity and density-dependent IMF (Eq.~\ref{eq:alpha3}) below and evolve them in such a way that the power-law index of the  present-day mass function at the low-mass end, i.e. $\alpha_1$ of the \citet{Kroupa2001} mass function, matches its predicted value from equation~13 of \citet{Baumgardt2003}. That is, the original SSP models without dynamical evolution are discarded and replaced by SSP models at the inferred age of the cluster and for the $\alpha_1$ value appropriate for this age. We then calculate the $M/L$ ratio which corresponds to this mass function. In the case of a top-heavy IMF, which is supported by several observational and theoretical studies (\citealt{Marks2012, Dabringhausen2009, Dabringhausen2010, Dabringhausen2012, Kroupa2013}), the inferred IMF slope in a sample of GGCs for stars with mass $M>1.0\Msun$ (i.e. $\alpha_3$) is flatter in more massive and denser environments \citep{Marks2012}. The variation of $\alpha_3$ can been described as follows,
\begin{equation}\label{eq:alpha3}
\alpha_{3} = \twopartdef {+2.3,}{x<-0.87,}{-0.41x+1.94,}{x\geq-0.87,}
\end{equation}
\noindent where $x$ is a function of $\FeH$ and birth GC cloud core density ($\rho_\mathrm{cl}$, stars plus gas) and is defined as follows,
\begin{equation}\label{eq:x}
x=-0.14 \FeH+0.99\log_{10}\left(\frac{\rho_\mathrm{cl}}{10^6\Msun\pc^{-3}}\right).
\end{equation}
According to \equref{eq:alpha3} the IMF becomes less top-heavy with increasing cluster metallicity and decreasing density. We refer to this IMF as the MKD IMF.

\begin{table*}[t!]
\centering
\caption{Comparison between the $M/L$ ratios of models and observed M31 GCs. The entries of the first column are respectively the adopted IMF, the AMR relation either from the Milky Way (MW) or M31, and the WD retention fraction. The entries of the second (third) column which are separated by a comma are respectively K-S statistic $D$ and the SSR divided by the number of data points in the $K$ ($V$) band. $\sigma$ is the standard deviation measured for each statistic by bootstrapping. The number of bootstrap-resamples is 10000. The first row corresponds to the best model of \citetalias{Hasani2016}. Smaller values of $D$ and SSR indicate a better agreement with the observation, i.e. the last two rows represent the best models.}\label{tab:KS-test}
\begin{tabular}{ccc}
\hline
\hline
Model & $K$ band & $V$ band\\
IMF + AMR relation + WD retention fraction & ($D\pm\sigma_{D}$, $SSR+\sigma_{SSR}$) & ($D\pm\sigma_{D}$, $SSR+\sigma_{SSR}$)\\
\hline
MKD IMF + No AMR + 50\% WD & $0.39 \pm 0.04$, \quad $0.78 \pm 0.11$ & $0.55 \pm 0.04$, \quad $2.69 \pm 0.28$ \\
canonical + AMR (MW) + 100\% WD & $0.18 \pm 0.03$, \quad $0.15 \pm 0.02$ & $0.20 \pm 0.03$, \quad $0.35 \pm 0.03$ \\
canonical + AMR (MW) + 50\% WD & $0.25 \pm 0.03$, \quad $0.16 \pm 0.02$ & $0.29 \pm 0.04$, \quad $0.32 \pm 0.03$ \\
canonical IMF + AMR (M31) + 100\% WD & $0.39 \pm 0.04$, \quad $0.20 \pm 0.02$ & $0.36 \pm 0.04$, \quad $0.33 \pm 0.04$ \\
canonical IMF + AMR (M31) + 50\% WD & $0.45 \pm 0.04$, \quad $0.22 \pm 0.02$ & $0.47 \pm 0.04$, \quad $0.37 \pm 0.04$ \\
MKD IMF + AMR (MW) + 100\% WD & $0.31 \pm 0.04$, \quad $0.44 \pm 0.07$ & $0.40 \pm 0.04$, \quad $1.22 \pm 0.12$ \\
MKD IMF + AMR (MW) + 50\% WD & $0.20 \pm 0.04$, \quad $0.31 \pm 0.05$ & $0.32 \pm 0.05$, \quad $0.81 \pm 0.09$ \\
MKD IMF + AMR (M31) + 100\% WD & $0.16 \pm 0.03$, \quad $0.20 \pm 0.04$ & $0.08 \pm 0.03$, \quad $0.40 \pm 0.06$ \\
MKD IMF + AMR (M31) + 50\% WD & $0.22 \pm 0.03$, \quad $0.19 \pm 0.03$ & $0.12 \pm 0.04$, \quad $0.34 \pm 0.04$ \\
\hline
\end{tabular}
\end{table*}

In our dynamical models we allow the initial mass of the GCs to vary and to depend on their current observed mass (taken from \citealt{Strader2011}) with a mean initial to final mass ratio $=3\pm1.5$. The initial half-mass radius of each GC is then determined using the relation between the initial mass and the half mass radius taken from \citet{Marks2010}. For the retention fraction of white dwarfs (WD), neutron stars (NSs) and black holes (BHs) we follow the same procedure as in \citetalias{Hasani2016}, i.e. the WD retention fraction is either 100\% or 50\% and the retention fraction of BHs and NSs depends on the cluster mass and radius and varies from $0\%$ (for the least massive and extended clusters) to 70\% (for the most massive and compact GCs). Further details and corresponding equations on how we take the dynamical evolution into account are given in \citetalias{Hasani2016} (sections 2 and 3 therein). 

To make a model cluster (shown as red crosses in \figref{fig:scatter-canonical}), the [Fe/H] value of an observed cluster is taken, and then a number of cluster masses are generated for the [Fe/H] value, each then being evolved to the age as given by the AMR. The red crosses in \figref{fig:scatter-canonical} show the $M/L$ ratios of our SSP models considering dynamical evolution and the AMR. The fact that the model data points are scattered in the $M/L-\FeH$ plane is due to their differences in the initial mass, radius and consequently the retention fraction. In this figure the stellar IMF in our model GCs is canonical. More interesting is the density and metallicity dependent IMF (Eq.~\ref{eq:alpha3}). As depicted in figures~\ref{fig:scatter-topheavy-MW} and \ref{fig:scatter-topheavy-And}, this IMF leads to better agreement with the observed trend and scatter in the $M/L$ ratios of M31 GCs. A comparison between our results in the present paper (Figures~\ref{fig:scatter-topheavy-MW} and \ref{fig:scatter-topheavy-And}) and our results in \citetalias{Hasani2016}, reveals that the incorporation of the AMR has improved the consistency of our results for a metallicity- and density-dependent top-heavy IMF, with the observed trend of M31 GCs.

To quantify the consistency between our results and observations, we apply the two-sample Kolmogorov-Smirnov (K-S) goodness of fit test, i.e. we determine the maximum difference, referred to as the K-S statistic $D$, between the empirical (cumulative) distribution function of the predicted $M/L$ ratios and that of the observed ratios of M31 GCs and compare it for different models. Moreover, we derive the sum of the squared residuals (SSR) between our models and those of M31 GCs. Smaller values of $D$ and SSR indicate a better agreement with the observation. \tabref{tab:KS-test} summarizes the K-S statistic $D$ as well as the SSR for different models, where the first row corresponds to our best model in \citetalias{Hasani2016}. We have also calculated the standard deviation ($\sigma$) of each statistic to quantify the significance of our statistics. The values of $\sigma$ are calculated through bootstrapping using 10000 bootstrap-resamples.

\par According to this table one can see that the models with the AMR show a far better agreement with observations in terms of the K-S statistic $D$ and SSR, especially if one uses the MKD IMF. This is also supported by comparing figures~\ref{fig:scatter-canonical} and \ref{fig:scatter-topheavy-And}. As one can see in \figref{fig:scatter-canonical}, the canonical IMF with the AMR systematically underestimates the $M/L$ ratios of the M31 GCs and does not reproduce the observed scatter in the $M/L-\FeH$ plane. However, the MKD IMF reproduces the observed trend and the scatter very well. Moreover, among the models with the MKD IMF, models with the AMR for  M31 have the lowest values of $D$ and SSR and their $\sigma$ values indicate that they are statistically better than models with the AMR for the MW. According to \tabref{tab:KS-test}, the models with canonical IMF + AMR (MW) do better than all other models except for models with MKD IMF + AMR (M31). In particular, in the $K$ band, the agreement of canonical IMF + AMR (MW) with the data of M31 GCs is comparable with that of MKD IMF + AMR (M31). However, in the $V$ band the MKD IMF + AMR (M31) does better (by $3\sigma$) in terms of the K-S statistic $D$. As a result, one can conclude that models with with MKD IMF + AMR (M31) provide the best agreement with the observed GCs in M31.

\par In terms of the WD retention fraction, one can see that models with smaller retention fractions are slightly better (except for the canonical IMF); however, the difference is not as significant as the difference between the adopted AMRs. The fact that lower WD retention fractions are better can be explained by the higher $M/L$ ratios of WDs compared to main-sequence stars. As a result, by decreasing the number of WDs in a cluster, the total $M/L$ ratio of the cluster decreases. Since the SSP models (\figref{fig:SSP}) have a larger $M/L$ ratio at the high-metallicity end compared to M31 GCs, a lower WD retention fraction helps to reduce the $M/L$ ratio of the models making them lie closer to the observed data. For the canonical IMF, a lower WD retention fraction is actually not beneficial. This is due to the fact that in the models with the canonical IMF and the AMR, the $M/L$ ratio is systematically below the average $M/L$ ratios of M31 GCs and reducing the WD retention fraction pushes the $M/L$ ratios of our models even further down.


\section{Summary and Conclusion}\label{sec:conclusion}
We studied the consequences of the AMR on the present-day $M/L$ ratios of GCs using dynamical SSP models of GCs. We considered two different cases of a canonical and a density and metallicity dependent (MKD) IMF. We demonstrated that the AMR leads to an anti-correlation in the observed $M/L$ ratios of GCs with respect to metallicity which is present for both cases of the adopted IMF.

\par In \citetalias{Hasani2016} we showed that SSP models with the MKD IMF and standard dynamical evolution can successfully explain the observed $M/L$ trend of M31 GCs. The present paper complements our previous work and shows that by taking the AMR into account, we can reach an even better agreement with observations than what we achieved in \citetalias{Hasani2016}. We demonstrated that regardless of the assumed IMF (canonical or density and metallicity dependent), models with the AMR are always better. Moreover, we showed that the AMR of M31 provides a better description of the observed M31 data compared to the AMR of the MW.

\par Our proposed solution to explain the observed $M/L$ ratios of M31 GCs is preferred over studies such as \citet{Shanahan2015} which use the canonical IMF. \citet{Shanahan2015} were able to explain the lower than expected $M/L$ ratios of M31 GCs. However, they were not able to explain the observed trend and the scatter in the $M/L-\FeH$ plane. More precisely, there are a number of metal-poor GCs in M31 which have a higher $M/L$ ratio than SSP models with a canonical IMF (\figref{fig:SSP}). The solution proposed by \citet{Shanahan2015} is not able to explain these data points. One mechanism to reproduce these data points, as we showed in the present paper, is by using the MKD IMF which makes more remnants than a canonical IMF and therefore leads to a higher $M/L$ ratio at the low metallicity end, and consequently a better agreement with the observations.

\par As a result, the combination of the AMR, a metallicity- and density-dependent top-heavy IMF and standard dynamical evolution driven by two-body relaxation, can be considered as a promising solution to the $M/L-\FeH$ problem of M31 GCs. As a final remark, it is encouraging that the IMF variation with metallicity and density of \equref{eq:alpha3} (MKD IMF), which was inferred from completely different data, leads to such good agreement with the M31 GC data studied here.


\acknowledgments
We would like to thank the anonymous referee for their detailed and helpful comments which improved the quality of this work.


\bibliographystyle{aasjournal}
\bibliography{manuscript.bbl}


\end{document}